\documentstyle[12pt]{article}
\begin{document}

\begin{titlepage}

\pagestyle{empty}

\begin{flushright}
{\footnotesize Brown-HET-992

April 1995}
\end{flushright}

\vskip 1.0cm

\begin {center}
{\large \bf Thermodynamic Properties of $\gamma$--Fluids
and the Quantum Vacuum}

\vskip 1.0cm

{J. A. S. Lima$^{1, 2}$ and A. Maia Jr.$^3$}

\end{center}

\vskip 0.5cm

\begin{quote}
{\small $^1$ Physics Department, Brown University\\
Providence, RI 02912, USA

$^2$ Departamento de F\'{\i}sica Te\'orica e Experimental\\
     Universidade Federal do Rio Grande do Norte\\
     59072-970, Natal - RN - Brazil

$^3$ Instituto de Matem\'atica - UNICAMP\\
13081 - 970 Campinas - SP - Brazil}
\end{quote}

\vskip 2.5cm

\begin{abstract}
\noindent The thermodynamic behaviour of a relativistic perfect simple fluid
obeying the
equation of state $p=(\gamma-1)\rho $, where
$0 \le \gamma \le 2$ is a constant, has                   been investigated.
Particular cases include: vacuum($p=-\rho $, $\gamma=0$), a randomly oriented
distribution
of cosmic strings ($p=-{1 \over 3} \rho $, $\gamma =2/3$),
blackbody radiation ($p={1\over 3}
\rho$, $\gamma =4/3$) and stiff matter ($p=\rho$, $\gamma=2$). Fluids with
$\gamma <1$ become hotter when they
expand adiabatically ($T\propto V^{1- \gamma}$). By assuming that such fluids
may be regarded as a kind of generalized radiation, the general
Planck's type form of the
spectrum is deduced. As a limiting case, a new Lorentz invariant spectrum of
the
vacuum which is compatible with the equation of state and  other
thermodynamic constraints is proposed. Some possible consequences to the early
universe physics are also discussed.
\end{abstract}

\end{titlepage}

\pagebreak

\section{Introduction}

\hspace{.3in} The concept of vacuum has pervaded the development of our
understanding about
space, matter and forces in the universe since the ancient
greek philosophers
\cite{filos}. In the same way that quantum mechanics was a
major breakthrought
for the theories of ordinary matter, so it was for modern
physical models of
vacuum. The first advance arose already in the years of the
old quantum
theory. It is closely related to the possible existence of a
zeropoint energy
for the blackbody radiation. In fact, the random background
radiation
corresponding to the zeropoint field is, presently, the key ingredient
of the so called
Stochastic Electrodynamics (SED) \cite{pena}-\cite{Boyer2}.
With the development of the
Quantum Eletrodynamics (QED) and other quantum field theories a new
concept arose,
namely, the physical vacuum is the ground state of a system
of quantum fields on the space-time manifold.
But now, we have to address at least two problems: firstly,
how to single
out the vacuum state? Secondly, what is an
intuitive picture of the
physical vacuum?

The answer for the first question depends on the
quantization method
used, as well as on the observer. For instance, in canonical quantization the
vacuum state is defined as
that one which contains no quanta. Technically, this means
that the effect of
annihilation operator acting on the state gives zero. On the other hand, when
using
functional methods in
quantum field theory, the vacuum state is defined as the
state which realizes
the minimum of the so called Effective Potential
\cite{coleman}. Recently,
some  authors addressed the issue that both of the above definitions could give
different answers
for the vacuum energy density \cite{myers}. Such drawbacks
are  present even
for Quantum Field Theories (QFT) formulated in the Minkowski
spacetime when a particle
detector is
uniformly accelerated \cite{citado}.

For the second question, two different approaches may be found
in the literature. Stochastic Electrodynamics
postulates the vacuum
as a random background of real electromagnetic fields endowed
with a well
defined frequency spectrum ($\rho (\nu) \propto \nu ^{3}$),
whereas for QED
the vacuum is filled with so called virtual pairs of
particles
(electron -- positron pair) whose direct detection is not
possible. Usually, as a kind of paradigm, QFT takes this last picture for
granted.

In the sixties, it was remarked that Lorentz invariance of
the vacuum requires
an energy momentum tensor (EMT) of  the form \cite{gliner}
\begin{equation} \label{eq:EMT}
<T_{\mu \nu}> = <\rho> \eta _{\mu \nu} ,
\end{equation}
where $\rho $ is the energy density and $\eta _{\mu \nu }$ is
the Minkowski
tensor. Therefore, the EMT of the vacuum describes a
particular relativistic
perfect simple fluid for which the equation of state is $p=
- \rho $ (see Eq. (\ref{eq:TAB})
bellow). In addition, since the energy momentum tensor
(\ref{eq:EMT})
 is divergenceless,
the vacuum energy density is constant in space-time. Besides,
performing a change of
inertial frame, the energy density of a fluid transforms as
\cite{weinberg}:
\begin{equation} \label{eq:ROL}
   \rho ' = {\rho + p {v^{2} \over c^{2}} \over 1 - {v^{2}
\over c^{2}}} ,
\end{equation}
where $v$ is the relative velocity between the frames. Thus,
it follows from
the equation of state that the energy density of the vacuum
is a Lorentz
invariant quantity, regardless of the form of its frequency
spectrum. In other
words, all inertial observers are comoving with the vacuum
background.

In this work we are mainly interested in the above
macroscopic point of view.
It will be assumed that the vacuum state of any bosonic or
fermionic field is
the less rigid state of matter compatible with the relativity
theory
\cite{vsom}. As we will  see, regarding the vacuum as an
unusual
substance described by the equation  $p=-\rho $ , the overall
thermodynamic properties of
it can be easily deduced. As in the case of blackbody
radiation, such
properties shed light on the underlying nature of the quantum
vacuum, determining,
for instance, the general form of its frequency spectrum. For the
sake of generality and also to simplify the comparison between the
vacuum and the blackbody radiation properties, we will consider a
monoparametric class of $\gamma$-fluids for which radiation and vacuum
are two important particular cases.

	The paper is organized as follows: In section 2 we present the
general thermodynamical properties of a $\gamma $--fluid. In section 3
we deduce the spectrum of a $\gamma $--fluid assuming as a natural ansatz that
such spectrum is a Wien's type. In section 4 we present a formal deduction
of such spectrum without assuming any ansatz while, in section 5,
we specialize to
the case ($\gamma =0$) and some striking consequences are founded for
the vacuum case, which are opposite to what happens for ordinary matter. In
section
6, Einstein's derivation of the blackbody radiation spectrum is generalized in
order to include the family of $\gamma $--fluids. In section
7, some consequences of our approach to early universe physics
are discussed. Finally, in section 8 we conclude with some
comments.

\section{Thermodynamics of a $\gamma$--Fluid}

\hspace{.3in} The thermodynamic states of a relativistic simple fluid are
characterized by
an energy momentum tensor $T^{\alpha \beta}$, a particle current $N^{\alpha }$
and an entropy
current $S^{\alpha}$. For a perfect fluid such quantities are
defined by \cite{dixon}
\begin{equation}
       T^{\alpha \beta}=(\rho + p)u^{\alpha} u^{\beta} -
	pg^{\alpha \beta},
        \label{eq:TAB}
\end{equation}
\begin{equation} \label{eq:NA}
        N^{\alpha}=nu^{\alpha},
\end{equation}
\begin{equation} \label{eq:SA}
        S^{\alpha}=n\sigma u^{\alpha},
\end{equation}
where $\rho$ is the energy density, $p$ is the pressure, $n$
is the particle number
density and $\sigma$ is the specific entropy (per particle).
The quantities
$p$, $\rho$, $n$ and $\sigma$ are related with the temperature
by the Gibbs law
\begin{equation} \label{eq:GIBBS}
          nTd\sigma = d\rho - {\rho + p \over n}dn \ .
\end{equation}

The basic quantities are constrained by the following
relations:
\begin{equation} \label{eq:CTAB}
        T^{\alpha \beta};_{\beta}=0 , \end{equation}
\begin{equation} \label{eq:CNA}
        N^{\alpha};_{\alpha}=0 ,
\end{equation}
\begin{equation} \label{eq:CSA}
        S^{\alpha};_{\alpha}=0 ,
\end{equation}
where semicolon denotes covariant derivative.

Equations (\ref{eq:CTAB}) and (\ref{eq:CNA}) express,
respectively the energy--momentum and the number
of particles conservation laws whereas
(\ref{eq:CSA}) is the thermodynamic
second law restricted to an adiabatic flow (``equation of
continuity'' for
entropy).

Bearing in mind the applications discussed ahead, first,
the temperature evolution equation will be obtained. From (\ref{eq:CTAB}) and
(\ref{eq:CNA}) it follows that
\begin{equation}\label{eq:rodot}
    \dot \rho + (\rho + p)\theta = 0 ,
\end{equation}
\begin{equation}\label{eq:ndot}
    \dot n + n \theta = 0,
\end{equation}
where an overdot means comoving  time derivative (for
instance, $\dot \rho
= u^{\alpha }\rho _{;\alpha }$) and $\theta = u^{\alpha
}_{;\alpha }$ is the
scalar of expansion.  Further, using (\ref{eq:ndot}) and
taking  $n$ and $T$ as independent
thermodynamic variables, Eq. (\ref{eq:rodot}) can be rewritten
as
\begin{equation} \label{eq:rot}
    \biggl({\partial \rho \over \partial T}\biggr)_{_n}\dot T =
     [n\biggl({\partial \rho \over \partial n}\biggr)_{_T} - \rho -p]
\theta \ \ \ ,
\end{equation}
and since $d\sigma $ is an exact differential,
the Gibbs law
(\ref{eq:GIBBS}) yields the well known thermodynamic
identity
\begin{equation}\label{eq:tro}
    T\biggl({\partial p \over \partial T}\biggr)_{_n}=
     \rho + p - n \biggl({\partial \rho \over \partial n}\biggr)_{_T}.
\end{equation}

Finally, replacing (\ref{eq:tro}) into (\ref{eq:rot}) and
using  (\ref{eq:ndot}) again one has \begin{equation}
\label{eq:EVOLT}
       { \dot T \over T} = \biggl({\partial p \over \partial
\rho}\biggr)_{_n}
                         {\dot n \over n}.
\end{equation}

In what follows, we consider the class of fluids described by the
``gamma law'' equation of state:
\begin{equation} \label{eq:GAMALAW}
        p=(\gamma - 1) \rho ,
\end{equation}
where the ``adiabatic index'' $\gamma$ lies on the interval $[0,2].$
This generalized equation of state accounts for a one-parametric family
of fluid systems, including a subclass with negative pressure. The limit
cases of vacuum $(\gamma=0)$ and stiff matter$(\gamma=2)$ are determined from
causality requirements
\cite{vsom}, whereas $\gamma = \frac{4}{3} $ describe the photon fluid.
With this choice, a straightforward
integration of (\ref{eq:EVOLT})
furnishes
\begin{equation} \label{eq:NT}
        Tn^{1- \gamma}=const ,
\end{equation}
and since $n$ scales with $V^{-1}$, where $V$ is the volume
of the considered
portion within the fluid, Eq. (\ref{eq:NT}) assumes the form
(from now on only the case $\gamma \neq 1$ will be considered)
\begin{equation} \label{eq:TV}
        T^{1 \over \gamma - 1 }V=const ,
\end{equation}
which is the usual adiabatic law for fluids with conserved
net number of
particles \cite{calvao}.

One additional thermodynamic constraint obeyed by the $\gamma$--fluid
is the generalized Stefan--Boltzmann law, namely:
\begin{equation}\label{ROTETA}
	\rho (T)=\eta T^{{\gamma \over \gamma - 1}},
\end{equation}
where $\eta$ is a $\gamma$--dependent constant. As shown in Ref.
\cite{janilo}, the above expression may be derived by at least,
three different methods among
them that one applied by Boltzmann (Carnot Cycle). It also follows
naturally from adiabatic condition
($dS=0$) when one considers the $\gamma$--law equation of state. Note that for
$\gamma =4/3$ it reduces to $\rho =\eta T^{4}$ and for the vacuum
case $(\gamma=0)$ it reduces to $\rho$=const as one should expect. As we
shall see, the interesting point here is that the relation (\ref{ROTETA})
together with Eq.(\ref{eq:NT}) rewritten as
\begin{equation}\label{NTNEW}
	n(T)=const \, T^{1 \over \gamma -1},
\end{equation}
are the thermodynamic constraints fixing the general form of the spectrum for
a $\gamma$--fluid, including, of course, the vacuum spectrum itself.

\section{The $\gamma$--Fluid Spectrum}

\hspace{.3in} In order to see how the thermodynamic constraints work to fix the
general
form of the spectrum, we make, initially, a simple and natural ansatz, which
is nothing more than a generalization of Wien's law obeyed by the
radiation fluid. In fact, the approach used below was earlier applied
to the case $\gamma=0$ in Ref.\cite{lima}.
In the next section a formal deduction of such a spectrum will be presented.

Firstly, it will be assumed that the $\gamma$--fluid is a kind of radiation
(like blackbody radiation), whose spectrum is Wien's type one, that is,
\begin{equation} \label{eq:WIEN}
       \rho _{T}(\nu ) = const \, \nu ^{\beta} \phi (\nu
	^{\beta}T^{\lambda}) ,
\end{equation}
where $\phi $ is an arbitrary function  and
$\beta $ and $\lambda $ are constants to be determined by using the constraint
equations (\ref{ROTETA}) and (\ref{NTNEW}), namely:
\begin{equation} \label{eq:rOCONST}
         \rho (T) =\int_{0}^{\infty} \rho_{T}(\nu )d\nu =
		 const \, T^{\gamma \over \gamma -1} \ ,
\end{equation}
and
\begin{equation} \label{eq:nCONST}
          n(T) = \int_{0}^{\infty} {\rho_{T}(\nu) \over h\nu}d\nu
          =const \,  T^{-{1 \over 1-\gamma}}.
\end{equation}

Substituting the spectrum (\ref{eq:WIEN}) into the constraint equations and
defining a new variable $u= \nu ^{\beta}T^{\lambda}$,
we can write
\begin{equation} \label{eq:ROF}
        \rho (T) = T^{-{\lambda (1+ \beta) \over \beta }}
		\int_{0}^{\infty} f(u)du =
		const \, T^{\gamma \over \gamma -1}
\end{equation}
and
\begin{equation} \label{eq:NG}
        n(T)=  T^{- \lambda} \int_{0}^{\infty} g(u)du = const \, T^{1 \over
		\gamma -1},
\end{equation}
where $f(u)$ and $g(u)$ are functions related with the original arbitrary
function  $\phi (u)$.

Comparing the powers of $T$ in Eqs.  (\ref{eq:ROF}) and
(\ref{eq:NG}) we get
\begin{equation}\label{lambeta}
	\lambda = {1 \over 1 - \gamma}= -\beta .
\end{equation}

Further, replacing (\ref{lambeta}) into (\ref{eq:WIEN}) we obtain the spectral
function for the $\gamma$-fluid
\begin{equation} \label{eq:WIENTYPE}
        \rho _{T} (\nu) = \alpha \, \nu ^{1 \over \gamma -1} \phi
	({T \over \nu}),
\end{equation}
where $\alpha$ is a dimensional constant and $\phi (T/ \nu )$ an arbitrary
function of its argument. As expected, if $\gamma= 4/3$, Eq.(26) reduces to
the well known Wien's law for blackbody radiation. Note also that,
although assuming a Wien's type spectrum  (Eq.(\ref{eq:WIEN})) for
the $\gamma$-fluid, this is not a completely arbitrary assumption. We have
simply
assumed a more general law, in analogy with Wien's, which satisfies
the constraints (\ref{ROTETA}) and (\ref{NTNEW}) for $\gamma =4/3$.

\section{Wien's Type Law for $\gamma$--Fluids: A formal deduction}

\hspace{.3in} As a preliminary point of principle, we recall
that if a hollow cavity
containing blackbody radiation changes its volume, adiabatically,
the ratio
between the energy and the corresponding frequency of each
component remains
constant, namely:
\begin{equation}\label{ENI}
	{E_{\nu}\over \nu }=const \, ,
\end{equation}
for any ``proper oscillation''.

This result, usually called the theorem of adiabatic invariance, holds indeed
for an arbitrary oscillating system when one of its parameter is slowly
modified by some external effect. For blackbody radiation the constancy of
the above quantity also means that for each band the mean number of quanta is
unaltered by reflection from the moving walls. In what follows, since the
overall existence of this adiabatic invariant can be proved regardless of the
nature of the oscillating system (see for instance,
Ref.\cite{ehrenfest}), its
validity will be assumed for the whole family of radiative
$\gamma$--fluid with $\gamma \ne 1$.

Now, if $\rho _{T} (\nu)$ is the spectral energy density inside an
enclosure with volume $V$, Eq.(\ref{ENI}) may be rewritten as
\begin{equation}\label{ROV}
	{\rho _{T} (\nu) d\nu V \over \nu} = const.
\end{equation}

	Note also that due to the thermal equilibrium state,
the energy density
in the band $d\nu $ varies with the temperature in the same manner as the
total energy density (in principle, only this band could
be present in the
cavity). Hence from the generalized Stefan--Boltzmann law, the above
adiabatic invariant takes the form
\begin{equation}\label{TVNU}
		{T^{\gamma \over \gamma -1} V \over \nu} = const ,
\end{equation}
and since $T^{1 \over \gamma - 1}V=const$ (see Eq.(\ref{eq:TV})), it follows
that $T/\nu $ is invariant. Thus, whether one compress or expand
adiabatically a hollow cavity containing a radiation $\gamma$--fluid, then
\begin{equation}\label{LT}
		\lambda T=const .
\end{equation}
The above result means that the displacement Wien's law, which is valid for
photons ($\gamma = 4/3$), holds in fact for the entire one--parametric family
of
$\gamma$--fluids.

Before discussing the remarkable physical consequences of Eq.(\ref{LT}) on
the vacuum state, we proceed to determine its effects on the general form of
the spectral distribution. To that end, we consider an enclosure containing
$\gamma $--fluid at temperature $T_{1}$ and focus our attention on the band
$\Delta \lambda _{1}$ centered on the wavelength $\lambda _{1}$ whose energy
density is $\rho _{T_{1}}(\lambda _{1})\Delta \lambda _{1}$.

If the temperature $T_{1}$ changes to $T_{2}$ due to an adiabatic expansion
(Note that $T_{1}$ does not necessarily decrease), the energy of the band
changes to
$\rho _{T_{2}}(\lambda _{2})\Delta \lambda _{2}$ and according to
Eq.(\ref{LT}) $\Delta \lambda _{1}$ and $\Delta \lambda _{2}$ are related
by
\begin{equation}\label{DELTL}
	{\Delta \lambda _{2} \over \Delta \lambda _{1}} = {T_{1} \over T_{2}}.
\end{equation}

Now, since one can assume that distinct bands do not interact, it follows that
\begin{equation}\label{RODELT}
	{\rho _{T_{2}}(\lambda _{2})\Delta \lambda _{2} \over \rho
	_{T_{1}}(\lambda _{1})\Delta \lambda _{1}}  = ({T_{2} \over
	T_{1}})^{\gamma \over \gamma - 1} .
\end{equation}

By combining the above result with (\ref{DELTL}) we conclude that
\begin{equation}\label{ROT}
	{\rho _{T_{2}}(\lambda _{2}) \over \rho
	_{T_{1}}(\lambda _{1})}  = ({T_{2} \over
	T_{1}})^{2\gamma - 1 \over \gamma - 1} ,
\end{equation}
and using again the displacement law given by (\ref{LT}),
we obtain for an arbitrary component
\begin{equation}\label{ROLg}
	\rho_{T} (\lambda ) \lambda ^{{2\gamma -1 \over \gamma -1}} = const.
\end{equation}

Note that in the case of blackbody radiation the above expression reduces to
$\rho _{T} (\lambda )\lambda ^{5} = const$, as it should be. Of course, due
to Eq.(\ref{LT}) the above result takes the form
\begin{equation}\label{ROLgN}
	\rho_{T} (\lambda ) = const \, \lambda ^{1-2\gamma \over \gamma -1}
	\phi (\lambda T),
\end{equation}
where $\phi $ is an arbitrary function of its arguments. Now in terms of the
frequency, since $\rho _{T}(\nu )d\nu =  \rho _{T}(\lambda )\mid {d\nu
\over d\lambda } \mid d\lambda $ it is easy to see that (\ref{ROLgN})
can be rewritten as
\begin{equation}\label{RONgN}
	\rho_{T} (\nu ) = \alpha \nu ^{1 \over \gamma -1}\phi ({T\over \nu }),
\end{equation}
as obtained in the previous section (see Eq.(\ref{eq:WIENTYPE})).

\section{Thermodynamics and the Vacuum Spectrum}

\hspace{.3in} In the case of blackbody radiation ($\gamma = 4/3$)
Eq.(\ref{eq:TV}) reduces
to $T^{3}V = const$, a well known result, while for the vacuum state ($\gamma
=0$) we obtain
\begin{equation}\label{eq:TVRAD}
	T=const \, V .
\end{equation}

We have therefore reached the conclusion that the vacuum becomes hotter if it
undergoes an adiabatic expansion. Such a result may be compared with those ones
of the usual theory of fluids for which $\gamma >1$ ($p>0$).

It should be emphasized that in the derivation of (\ref{eq:TV}) the
conservation of the number of particles was explicitly used. However, the
meaning of such an assumption needs to be clarified. For $p={1\over 3} \rho $
we see from  (\ref{eq:NT}) that $n$ scales
with $T^{3}$. As we know, since the chemical potential
of photons is zero its total number is indefinite
so that $n$ must be
interpreted as the average number density of photons.
As is well known, such an interpretation is in agreement
with the Planck distribution which furnishes $n= \int_{0}^{\infty}
{\rho_{T}(\nu) \over h\nu}d\nu =bT^{3}$, where $b$ is a constant
\cite{landau}.

In what follows we assume that similar considerations hold for the vacuum
state ($\gamma =0$), for which Eq.(\ref{eq:NT}) yields
\begin{equation}\label{NCONST}
	n={const \over T}.
\end{equation}

Hence, we see that for the vacuum state, the average density of particles
decreases with increasing $T$. In the limit $T \rightarrow \infty $, $n$
goes to zero, being infinite in the opposite extremum ($T=0$). It
should be noticed that both results are consistent with
Eq.(\ref{eq:TVRAD}).

Let us now consider the vacuum spectrum itself. From the above results we can
say
that the energy spectrum $\rho _{T}(\nu )$ must satisfy two thermodynamic
constraints:
\begin{equation} \label{eq:roCONST}
        \rho = \int_{0}^{\infty} \rho_{T}(\nu )d\nu = const \, ,
\end{equation}
and
\begin{equation} \label{eq:nnCONST}
         n=\int_{0}^{\infty} {\rho_{T}(\nu) \over h\nu}d\nu = {const \over T},
\end{equation}
which are just the constraints (\ref{ROTETA}) and (\ref{NTNEW}) for
$\gamma =0$.

Taking $\gamma =0$ in Eq.(\ref{RONgN}), instead of the result
$\rho_{T}(\nu )=const \, \nu ^{3} $, claimed by the proponents of SED, we find
that the only Wien type spectrum for the vacuum state, compatible with the
thermodynamic constraints is given by
\begin{equation}\label{ROTFIN}
	\rho_{T} (\nu ) = const \, \nu ^{-1}\phi ({T\over \nu })
\end{equation}
a result obtained earlier in Ref. \cite{lima}.

It should be noticed that even in the limit $T \rightarrow
0$, $\rho_{T} (\nu)$ scales
with $\nu ^{-1}$ instead of $\nu ^{3}$, as usually inferred
from the blackbody
radiation spectrum \cite{pena}-\cite{Boyer2}.
In fact, the later result follows from our Eq.(36) by choosing $T=0$ and
$\gamma=4/3$, since zeropoint
radiation in the context of SED
satisfies $p= {\rho \over3}
$. Of course, this kind of vacuum is rather
different from the one considered here. In the present
case, we remark that the existence of a temperature
dependent spectrum for the vacuum state is not
forbidden by the relativity
principle, as long as the vacuum fluid is described
by the
equation of state
$p=-\rho $.

Finally, we would like to stress some physical consequences of the
displacement Wien's law to the case of the vacuum state or, in general,
for $\gamma$--radiation fluids with $\gamma <1$.

First, it should be recalled that if a blackbody radiation fluid expands
adiabatically its temperature is lowered ($T\propto V^{-1/3}$) and since
$\lambda T=const$, the wavelength of each band increases, thereby lowering
the total energy density in accordance with the Stefan--Boltzmann law. This
is the typical behavior for fluids with $\gamma>1$. For
$\gamma <1$, however, the temperature grows if the fluid undergoes an
adiabatic expansion ($T^{1\over \gamma - 1}V=const$). The increase in
       temperature is accompanied by a decrease in each wavelength
$\lambda$ in accordance with Wien's law . The vacuum state behaves like
 a limiting case of this subclass, the one for which the energy density remains
constant in the course of expansion.

\section{Planck's Type Spectrum of $\gamma$--Radiation}

\hspace{.3in} In this section we will derive, up to a constant,
a formula giving the
spectral distribution for generalized $\gamma$--radiation. As a limiting case,
a new Lorentz invariant spectrum
for the vacuum state will
be presented. Our derivation
will be carried out through a slight modification of the arguments used by
Einstein \cite{einstein} in his original deduction of the Planck radiation
spectrum which was
based on Wien's law plus some additional hypotheses concerning the
interaction between radiation and matter.

	Let us consider an atomic or molecular gas, the particles of which can
exist in a number of discrete energy levels $E_{n}=1,2,...$etc, in thermal
equilibrium with the $\gamma$--radiation at temperature $T$. The probability
that an atom is in the energy level $E_{n}$ is given by the Boltzmann factor
\begin{equation}\label{wn}
	W_{n}=p_{n}e^{-{E_{n}\over kT}},
\end{equation}
where $p_{n}$, the statistical weight of the nth quantum state, is
independent of the temperature.

	Of course, transitions happen by emission or absorption of
quanta of the
$\gamma$--radiation which satisfies the following hypotheses:

H$_{1}$) The $\gamma$--radiation spectrum is Wien's type, as deduced earlier,
namely:
\begin{equation}\label{alfani}
	\rho _{T}(\nu )=\alpha \nu ^{1 \over \gamma -1}\phi ({T\over\nu})
\end{equation}
where $\alpha $ is a dimensional $\gamma$-dependent constant.

H$_{2}$) Bohr's postulate for atomic emission or absorption remains valid for
quanta of  $\gamma$--radiation, that is,
\begin{equation}\label{emen}
	E_{m} - E_{n}=h\nu .
\end{equation}

	Following  Einstein, in such a system there exist three types of transition
processes by which equilibrium is established. The first one is due to
absorption of $\gamma$--radiation, with the atom making an upward transition
from $E_{n}$ to the level $E_{m}$ according to the probability, per unit time
\begin{equation}\label{wnm}
	\dot W_{nm}=B^{m}_{n}\rho _{T}(\nu ).
\end{equation}
where $B^m_n$ is a constant characterizing the specific transition.

The second one is spontaneous emission, which happens in the absence of any
$\gamma$--radiation and is determined by the coefficient $A^{n}_{m}$,
and, finally, the stimulated emission characterized by $B^{n}_{m}$. For these
processes the net transition probability per unit time is
\begin{equation}\label{wmn}
	\dot W_{mn}= A^{n}_{m}+B^{n}_{m}\rho _{T}(\nu ) .
\end{equation}
Hence, from equation (\ref{wn}) the equilibrium condition can be written as
\begin{equation}\label{pnpm}
	p_{n}e^{-E_{n}/kT} B^{m}_{n}\rho _{T}(\nu )=p_{m}e^{-E_{m}/kT}
				(B^{n}_{m}\rho _{T}(\nu )+A^{n}_{m}) ,
\end{equation}
and solving for the energy density one obtains
\begin{equation}\label{forend}
	\rho _{T}(\nu ) = {{p_{m}\over p_{n}} {A^{n}_{m}\over B^{n}_{m}} \over
			 e^{E_{m}-E_{n}\over kT} - {p_{m}\over p_{n}}
			 {B^{n}_{m}\over B^{m}_{n}}} .
\end{equation}

       	Now, at very high temperatures, it will be assumed that stimulated
emission is much more probable than spontaneous emission so that (\ref{pnpm})
leads to
\begin{equation}\label{pb}
	p_{n} B^{m}_{n}= p_{m} B^{n}_{m} ,
\end{equation}
and using H$_{2}$, Eq.(\ref{forend}) can be recast in the form
\begin{equation}\label{forend2}
	\rho _{T}(\nu )= { A^{n}_{m}/ B^{n}_{m} \over e^{h\nu /kT}-1 }.
\end{equation}
Finally, comparing (\ref{forend2}) with (\ref{alfani}) it follows that
\begin{equation}\label{abalfani}
	 {A^{n}_{m}\over B^{n}_{m}}=\alpha \nu ^{1\over \gamma -1} ,
\end{equation}
and
\begin{equation}\label{finit}
	\phi ({T\over\nu})={1\over e^{h\nu/kT}-1} ,
\end{equation}
with (\ref{forend2}) taking the form
\begin{equation} \label{eq:forend3}
	\rho _{T}(\nu )={\alpha \nu ^{1\over \gamma -1} \over e^{h\nu /kT}-1}.
\end{equation}
This is the most natural generalization of Planck's radiation formula for
$\gamma$--radiation \cite {rem}. Einstein's result follows for $\gamma =4/3$.
However, more interesting for fundamental physics, is that the spectrum for
the ``hot vacuum'' ($\gamma =0$) is given by
\begin{equation}\label{rovac}
	\rho ^{vac}_{T}(\nu )= {\alpha \nu^{-1} \over e^{h\nu/kT}-1}.
\end{equation}
{}From Eq.(\ref{eq:rOCONST}) or by straightforward integration of
the above equation, it is easy to see that
\begin{equation}\label{rovac2}
	\rho ^{vac}= \int _{0}^{\infty}\rho ^{vac}_{T}(\nu )d\nu = const ,
\end{equation}
as it should be.

It is worth mentioning that, in the
framework of QFT, several attempts have been made
to assign a definite spectrum to the vacuum
state in connection with the
so-called Casimir
effect \cite{ford}.
As we know, this effect is a response of the vacuum structure to constraints
imposed by spatial
boundaries or a nontrivial topology. As a matter
of fact, the Casimir spectrum and that one given above
are quite different, even considering that both do not have
a Planckian form. However, aside from some inevitable ambiguities
present in the former
(a fact explicitly recognized in the quoted papers),
we notice that (54) has been deduced as a direct consequence of the equation
of state $p= -\rho$. In particular, this means that
such a spectrum describes bulk properties
of the vacuum fluid instead of ``distortions"
effects produced by spatial boundaries as
occur, for instance, with blackbody
radiation in
microcavities or more generally,
with confined quantum gases \cite{actor}.
Naturally, since far from the boundaries
the vacuum properties must be the same as what it
would present in free space, one may expect
to recover (54) as a limiting case of a
proper Casimir spectrum. This issue is
presently under investigation.

	The above results show that, under
very reasonable hypotheses,
the spectrum of the family of $\gamma$-fluids
is uniquely determined up to
the constant scale factor $\alpha$. In principle,
the value of this constant could be determined
using the low frequences limit ($h\nu<<kT$).
However, unlike of the blackbody radiation case,
does not exist
presently an independent derivation of the
Rayleigh-Jeans(RJ) limit for arbitrary values
of the $\gamma$ parameter.
It should be noticed that for  h$\nu$
much smaller than $kT$, (\ref{eq:forend3})
leads to a RJ-type form given by
\begin{equation}\label{eq:Ray}
        \rho_{T}(\nu)=
         {\Omega kT \nu^{2-\gamma\over \gamma - 1}} \ \ \ ,
\end{equation}
where the unknown $\Omega$ is a dimensional $\gamma$--dependent function such
that
$\Omega({4\over 3})= {8\pi\over c^3}$. The question
related with the precise form of the RJ limit for a
$\gamma$--fluid and its influence on the fluctuations of energy  will be
discussed elsewhere.

        In terms of the wavelength, Eq.(\ref{eq:forend3}) may be
rewritten as
\begin{equation}\label{wave}
	\rho _{T}(\lambda) = {\beta \lambda^{1-2\gamma\over \gamma-1} \over
			 e^{hc\over k\lambda T} - 1} \ \ \ ,
\end{equation}
where $\beta$ is also a $\gamma$-dependent constant. As is well
known, for blackbody radiation the wavelength $\lambda_{m}$ for which
$\rho_T(\lambda)$ assumes its maximum value satisfies a
displacement law under the form \cite{landau}
\begin{equation}\label{degree}
	\lambda_{m} T=0.289 \ \ \mbox{cm.degree} \ \ .
\end{equation}

It turns out that the above result can be easily generalized for a
$\gamma$-fluid using Eq.(\ref{wave}). In fact, since $\lambda_m$ is
determined by the condition ${\partial \rho_T(\lambda)\over \partial
\lambda}=0$, it follows from (\ref{wave}) that
\begin{equation}\label{turns}
{\partial \over \partial \lambda} \biggl({\lambda^{1-2\gamma \over \gamma
-1} \over e^{hc \over kT\lambda} -1}\biggr) =0 \ \ \ ,
\end{equation}
or still
\begin{equation}\label{still}
x + b e^{-x} - b = 0
\end{equation}
where $x={hc \over k\lambda T}$ and $ b={2\gamma -1 \over \gamma -1}$.
Hence, if $p(\gamma)$ denotes the roots of the above equation, then
\begin{equation}\label{roots}
\lambda_m T = {hc \over kp(\gamma)} = {1.438 \over p(\gamma)}
\ \ \mbox{cm.degree}.
\end{equation}
Note that $x=0$ , or equivalently $p=0$, is a trivial solution of (\ref{still})
regardless of the value of $\gamma$. However, for $\gamma > 0$ there always
exists another physically meaningful solution. For instance,
if $\gamma = {4 \over 3}$ one has $p ({4 \over 3}) = 4.965$
so that the result (\ref{degree}) is recovered. Another example is provided
           by stiff matter$(\gamma=2)$ for which $p(2)= 2.821$ and from
(\ref{roots}),
$\lambda_{m}T = 0.510$ cm.
degree. For the vacuum$(\gamma=0)$ case, however, it is easy to show
that there only exists the trivial solution. In other words, the graph of
$ \rho_{T}(\lambda)$ does not exhibit a finite maximum value characteristic of
the class of $\gamma$-fluids. It is infinite for $\lambda=0$ and decreases
monotonically to zero when $\lambda $ goes
to infinity. As the reader may
conclude by himself, in fact, this is the only physical possibility allowed by
the Lorentz invariance of
the vacuum spectrum.

\section{Some Consequences in Cosmology}

\hspace{.3in} The above results may be interesting to early universe
physics mainly to
the so called inflationary models. The essential
feature of such
models is the appearance of an accelerated expansion
of the
universe driven by
the vacuum stress arising, for instance, from a scalar field
with a global
minimum in its effective potential or some types of phase
transition.
It turns out that negative pressure is the key condition
to generate either exponential or ``power law"inflation.

To the best of our knowledge the thermodynamic behavior of the
field driving inflation has so far been neglected. By
assuming that it
behaves like a perfect fluid with $p=(\gamma -1) \rho$, where
$\gamma < \frac{2}{3}$ for power law inflation, the results
presented here can be easily addapted. In fact, since our
results are generally
covariant, we can apply Eq. (\ref{eq:TV}) for a Friedman
Robertson-Walker metric
($V\propto R^{3}$) to obtain:
\begin{equation} \label{eq:RTFRW}
        T = T_{*}({R_{*} \over R})^{3(\gamma - 1)} ,
\end{equation}
where $R(t)$ is the universal scale function and $T_{*}=
T(R_{*})$ is the temperature when the scale factor takes    on the value
$R_{*}$. For $\gamma =
4/3$ one finds
$T\propto R^{-1}$ as usual for a radiation dominated
phase.
Special results
are: (i) Exponential inflation ($\gamma = 0 $, vacuum, $T
\propto R^{3}$), power-law inflation
($0<\gamma < \frac{2}{3}$,
$T\propto R^{3(1- \gamma)}$).

It should be noticed that the above result holds
regardless of the nature of the $\gamma$-fluid, that is, it
does not matter whether it
is regarded as a generalized radiation. As a matter of fact, the
temperature law given by (\ref{eq:RTFRW}) is a
consequence of the
``gamma-law" equation of state. For instance, it can be applied even
for dust $(\gamma=1)$ furnishing $T=T_*$=const. in accordance with
(\ref{eq:NT}). Another interesting example is
provided by a randomly oriented
distribution of infinitely thin straight strings
averaged over all directions.
As shown by Vilenkin \cite{vilenkin}, such a system
behaves like a perfect
fluid with $p={-1\over 3}\rho \ (\gamma ={2\over 3})$ and from
(\ref{eq:RTFRW}) we obtain $T=$ const $\times R$ \cite{janilo}.

An additional point, supporting the present treatment
of the vacuum as a fluid endowed with a definite temperature
dependent spectrum, appears at the interface of particle physics and
cosmology in the so-called cosmological constant
problem. In fact, as
is well known, the vacuum energy density due to the zeropoint
energy of all normal modes of a field contributes
to the
cosmological $\Lambda$-term which behaves
like a
fluid with
$p= - \rho$. However, the cosmological
estimates of such a term
$(\Lambda/_{8\pi G}$ \rlap{\raisebox{-.5ex}{$\scriptstyle{\sim}$}}
\hspace{-.2cm}{\raisebox{.3ex}{$<$} $10^{-47} Gev^4)$ is  smaller
than the estimates which follow from field theories
by at least, forty
orders of magnitude. This puzzle which makes up
the essence of the problem
has been the subject of numerous papers \cite{chen}-\cite{limao}.

A possible approach to circumvent this problem,
which has been
investigated in the recent literature
(see Ref.\cite{limao} and
references there in), is to assume that the effective
$\Lambda$-term is a fluid
interacting with the other matter fields of the universe (as in a
multifluid model). In this case, the vacuum energy
density is not
constant since the energy momentum tensor of
the mixture must be conserved in the course of the expansion.Thus, a
slow decaying of the vacuum energy density may
provide the source term for
material particles or radiation, thereby suggesting a natural
solution to this puzzle, namely: the cosmological
constant is very
small today because the universe is very old.

\section{Conclusion}

\hspace{.3in} In this paper we have attempted to give a systematic
treatment of how thermodynamics and semiclasical considerations can be used to
determine the spectrum
of a $\gamma$--fluid, including the vacuum spectrum as a particular case. The
physical motivation of such a study
is based on two different, although closely related features,namely:
the Lorentz invariance of the vacuum state
which requires that its energy-momentum tensor is
proportional to the Minkowski tensor, that is, a perfect fluid obeying the
equation of state $p= -\rho$
and the probable existence of an universal
$\Lambda$--term which is also equivalent,in the
cosmological domain, to a
vacuum fluid
satisfying the same equation of state.

For the sake of generality, several thermodynamic properties of $\gamma$-fluids
with
positive and negative pressures have been investigated.
In this connection, we remark that thermodynamic
states with negative pressures
are metastable but they are not forbidden by any law of nature. These
states are also hydrodynamically unstable for bubbles
and cavity formation and a spontaneous collapse
could also be expected \cite{landau}. As remarked in Ref.\cite{lima}
one may speculate whether such collapse may be
answerable
for the matter creation process from `` nothing"
with the particles being
ultimately described as a kind of vacuum condensation.

By regarding the class of $\gamma$-fluids as radiation
with
different equations of state, a formal deduction for
Wien's law has been presented and such a result allows
us to derive, up
to a dimensional constant, the generalized Planckian
type form of the
spectrum. Probably, only using QFT or statistical methods, such a
constant will be determined. It was also shown that in the limit of low
frequences the spectrum scales with
$kT\nu^{2-\gamma\over \gamma- 1}$. For comparison, the usual blackbody
expressions have sistematically been recovered by taking
$\gamma={4\over 3}$.
Further, as a special case, the
thermodynamic behaviour and the vacuum spectrum
satisfying the equation of
state $p=-\rho$ were obtained and its unusual features discussed in
detail. The vacuum temperature, or more generally, the temperature of
a $\gamma$-fluid  (for $\gamma <1$) increases in
the course of an
adiabatic expansion and, unlike blackbody radiation,
their wavelength
decreases as required  by Wien's law. In particular,
this explains
why the energy density of a pure vacuum (cosmological constant)
remains constant if a hollow cavity (universe)
undergoes an adiabatic
expansion.

We also argued that the unsettled situation arising
from the overall
existence of the vacuum and its consequences on the interface uniting
QFT, general relativity and cosmology may be
circumvented by a more
comprehensive picture of the vacuum itself.
Theoretically, as happens when one includes quantum
corrections to the general relativity, the
treatment of the vacuum as a fluid also suggests  a cosmological scenario
where the evolution may initially be supported by
a pure vacuum state.
By virtue of the expansion, the vacuum decays
generating all matter and
entropy existing in the universe, thereby explaining naturally the
small value of the $\Lambda$-term presently
observed ( See Ref. \cite{limao} for a cosmology
satisfying such conditions).

\section{Acknowledgments}

We would like to thank Professor R. Brandenberger
and Dr.
A. Sornborger for many valuable discussions
and for a critical reading of
the manuscript. One of us
(J.A.S.Lima)
is grateful for the
hospitality of the Physics
Department-Brown University
where part of this work was carried out. This paper
was
partially supported by the Conselho Nacional de
Desenvolvimento Cientifico--CNPQ (Brazilian Research
Agency).


\begin{thebibliography}{16}
\bibitem{rem} We notice that in the derivation
of (53),
hyphothesis H$_{2}$ does not play as prominent
a role as
it appears at first sight. One may
assume, for instance,
that in the limit of large frequences ($h\nu>>kT$)
Wien's law (36) is given by
$\rho _{T}(\nu)= \alpha \nu ^{1\over \gamma -1}
e^{-h\nu /kT}$ i.e.
$\phi ({T \over \nu}) = e^{-h\nu /kT}$ in
this limit. As a consequence, instead of
having the status of a fundamental hyphothesis,
the Bohr relation is
recovered as a consistency condition
regardless of the
value of $\gamma$. We avoid to
generalize this
historical derivation (see Ref.\cite{einstein}),
because there is no experimental
evidence for the
above law when $\gamma$ is different
from ${4 \over3}$.
\end{thebibliography}
\end{document}